\documentclass[twocolumn]{aastex701} 
\accepted{ApJL}
\usepackage{booktabs}
\usepackage{diagbox}
\usepackage{natbib}
\usepackage{amsmath} 
\usepackage{orcidlink}

\begin{document}

\title{Optical Light-Curve Broadening in the Redback Binary PSR J2215+5135}

\author{Amin Farhang*\,\orcidlink{0000-0001-7712-697X}}
\affiliation{School of Astronomy, Institute for Research in Fundamental Sciences (IPM), P.O. Box 19395-5531 Tehran, Iran}
\affiliation{Iranian National Observatory (INO), P.O. Box 19568-36613, Tehran, Iran}
\email[show]{a.farhang@ipm.ir}

\author{Soroush Shakeri\,\orcidlink{0000-0002-1765-0494}}
\affiliation{Department of Physics, Isfahan University of Technology (IUT), Isfahan 84156-83111, Iran}
\email{}

\author{Habib Khosroshahi\,\orcidlink{0000-0003-0558-8782}}
\affiliation{School of Astronomy, Institute for Research in Fundamental Sciences (IPM), P.O. Box 19395-5531 Tehran, Iran}
\affiliation{Iranian National Observatory (INO), P.O. Box 19568-36613, Tehran, Iran}
\email{}

\author{Narjes Shahamat\,\orcidlink{0009-0009-7526-4522}}
\affiliation{Department of Physics, Isfahan University of Technology (IUT), Isfahan 84156-83111, Iran}
\email{}

\author{Razieh Ranjbar\,\orcidlink{0000-0003-1488-4890}}
\affiliation{School of Astronomy, Institute for Research in Fundamental Sciences (IPM), P.O. Box 19395-5531 Tehran, Iran}
\email{}

\author{Jerome Orosz\,\orcidlink{0000-0001-9647-2886}}
\affiliation{ Astronomy Department, San Diego State University, San Diego, CA 92182-1221, USA}
\email{jorosz@sdsu.edu}

\author{Joshua Tan\,\orcidlink{0000-0003-3835-8115}}
\affiliation{Department of Astronomy, Columbia University, Mail Code 5246, 550 West 120th Street, New York, NY 10027, USA}
\email{}

\author{Davood Rafiei Karkevandi\,\orcidlink{0000-0002-6031-041X}}
\affiliation{Department of Physics, Isfahan University of Technology (IUT), Isfahan 84156-83111, Iran}
\email{}

\author{Hamed Altafi} 
\affiliation{Iranian National Observatory (INO), P.O. Box 19568-36613, Tehran, Iran}
\email{}

\author{Hamidreza Mahani\,\orcidlink{0009-0002-9957-5818}}
\affiliation{School of Astronomy, Institute for Research in Fundamental Sciences (IPM), P.O. Box 19395-5531 Tehran, Iran}
\email{}

\author{Amir Yarahmadi} 
\affiliation{School of Astronomy, Institute for Research in Fundamental Sciences (IPM), P.O. Box 19395-5531 Tehran, Iran}
\email{}

\author{Masoud Bidar} 
\affiliation{Iranian National Observatory (INO), P.O. Box 19568-36613, Tehran, Iran}
\email{}

\author{Ramin Shomali} 
\affiliation{Iranian National Observatory (INO), P.O. Box 19568-36613, Tehran, Iran}
\email{}

\author{Mohammadreza Rostamian} 
\affiliation{Iranian National Observatory (INO), P.O. Box 19568-36613, Tehran, Iran}
\email{}

\author{Hengameh Shalchian} 
\affiliation{Iranian National Observatory (INO), P.O. Box 19568-36613, Tehran, Iran}
\email{}

\author{Tayebeh Shokatpour} 
\affiliation{Iranian National Observatory (INO), P.O. Box 19568-36613, Tehran, Iran}
\email{}

\author{Ruholla Sadeghi Amin} 
\affiliation{Iranian National Observatory (INO), P.O. Box 19568-36613, Tehran, Iran}
\email{}

\begin{abstract}
Spider millisecond pulsars offer a direct route to constrain neutron-star masses, including candidates above two solar masses, via optical light-curve modeling of the irradiated companion. This assumes stable irradiation. We report new $g'r'i'$ photometry of the redback PSR~J2215+5135 from the Iranian National Observatory in 2024 and compare it with the 2014 data of \citet{linares2018peering}. Peak brightness and orbital period are unchanged within uncertainties, but the light-curve wings are significantly broader in 2024, as the local deviations exceed $3\sigma$ on the rising branch in all filters, with an additional $i'$-band excess on the falling branch, giving a global permutation-test significance of $4.26\sigma$. A single-hot-spot model fits the 2024 data better than a spot-free one, but systematic residuals remain. The broader wings are reproduced only through correlated shifts in the spot, inclination, filling factor and companion temperature, which we regard as model compensation, not physics. An extended, variable heating component may also fit. The best-fit spot temperature differs between epochs by about $1060$\,K, comparable to its model dependence (about $900$\,K between two codes for 2014), so we do not interpret this as physical. Two epochs cannot show whether the change is gradual, sudden or short-lived. The inferred neutron-star mass depends on the spot treatment: $2.13^{+0.16}_{-0.13}\,M_\odot$ for 2014 with a hot spot, and $M_\mathrm{NS}=1.95^{+0.11}_{-0.07}\,M_\odot$ for 2024. These agree within about $1\sigma$, but come from differently shaped light curves, and the statistical uncertainties exclude heating-prescription systematics.

\end{abstract}

\keywords{\uat{Millisecond pulsars}{1062} --- \uat{Neutron stars}{1108} --- \uat{Binary stars}{154} --- \uat{Stellar atmospheres}{1584} --- \uat{Stellar photometry}{1620} --- \uat{Binary stars}{154}}

\section{Introduction} 
Millisecond pulsars (MSPs) in compact binaries can be spun up via prolonged accretion of mass and angular momentum from a low-mass companion, a process known as recycling. A subclass of these systems, called `spider' pulsars \citep{1988Natur.333..237F}, hosts companions in tight orbits ($P_{\rm b} \lesssim 1$\,day) that are continuously irradiated by the pulsar wind. They are classified by companion mass into `black widows' ($M_c \lesssim 0.1\,M_\odot$) and `redbacks' ($M_c \sim 0.1$--$0.4\,M_\odot$). The irradiation heats the companion's day side and drives mass loss \citep{1988Natur.334..225K, 1988Natur.334..684V, 2013IAUS..291..127R}, producing large-amplitude orbital modulations in the optical light curve. Because the light-curve shape encodes the binary inclination, spider pulsars offer one of the few direct routes to neutron-star masses above $2\,M_\odot$ \citep{Romani2016, linares2018peering, 2020ApJ...892..101K}.

In a simple irradiation picture, the optical maximum should occur at superior conjunction of the companion. In practice, many spider light curves are asymmetric, with the brightness peak shifted in phase. These features point to a heating pattern that is not centered on the substellar point. Proposed explanations include illumination from an intrabinary shock (IBS) formed where the pulsar wind meets the companion outflow \citep{Romani2016}, magnetically channeled particle deposition near a companion magnetic pole \citep{Romani:2021xmb}, and atmospheric heat redistribution by surface winds \citep{Sullivan:2024qgl}. In several systems, adding a hot-spot component to the light-curve model substantially improves the fit, suggesting that the heating is concentrated in specific regions rather than distributed uniformly across the pulsar-facing hemisphere.

PSR~J2215+5135 is a redback MSP discovered in the \textit{Fermi}-LAT survey \citep{abdo2013second} and confirmed through radio timing \citep{hessels2011fermi}. It has been widely studied as a potential host of an exceptionally massive neutron star. Mass estimates from optical modeling span a wide range from $2.45^{+0.22}_{-0.11}\,M_\odot$ \citep{Schroeder:2014qwa}, $\sim\!1.6\,M_\odot$ \citep{romani2015massive}, $2.27^{+0.17}_{-0.15}\,M_\odot$ \citep{linares2018peering}, $2.28^{+0.10}_{-0.09}\,M_\odot$ \citep{2020ApJ...892..101K}, and $1.85$--$2.3\,M_\odot$ \citep{Sullivan:2024qgl}. This wide spread reflects the sensitivity of light-curve modeling to the assumed irradiation geometry and atmospheric prescription. PSR~J2215+5135 is relatively bright and shows smaller light-curve variability than several other redbacks, which is a main reason it is so often chosen for modeling and mass measurements. Redbacks are known to vary on timescales from minutes, through flares, to months, through substantial changes in surface heating \citep{2016ApJ...823..105D,2019ApJ...872...42S}, so the apparent stability of this source should not be assumed to hold at all times. This makes it a useful test case for whether surface heating changes over a decade.

Here we present new optical photometry of PSR~J2215+5135 obtained with the 3.4-m Iranian National Observatory (INO) in 2024, using the same SDSS $g'$, $r'$, and $i'$ filters as the 2014 William Herschel Telescope observations of \citet{linares2018peering}. The comparable apertures and filter sets enable a direct epoch-to-epoch comparison over a ten-year baseline.

\section{Observation \& Measurements} 
\begin{figure}[ht!]
\includegraphics[width=1.0\linewidth]{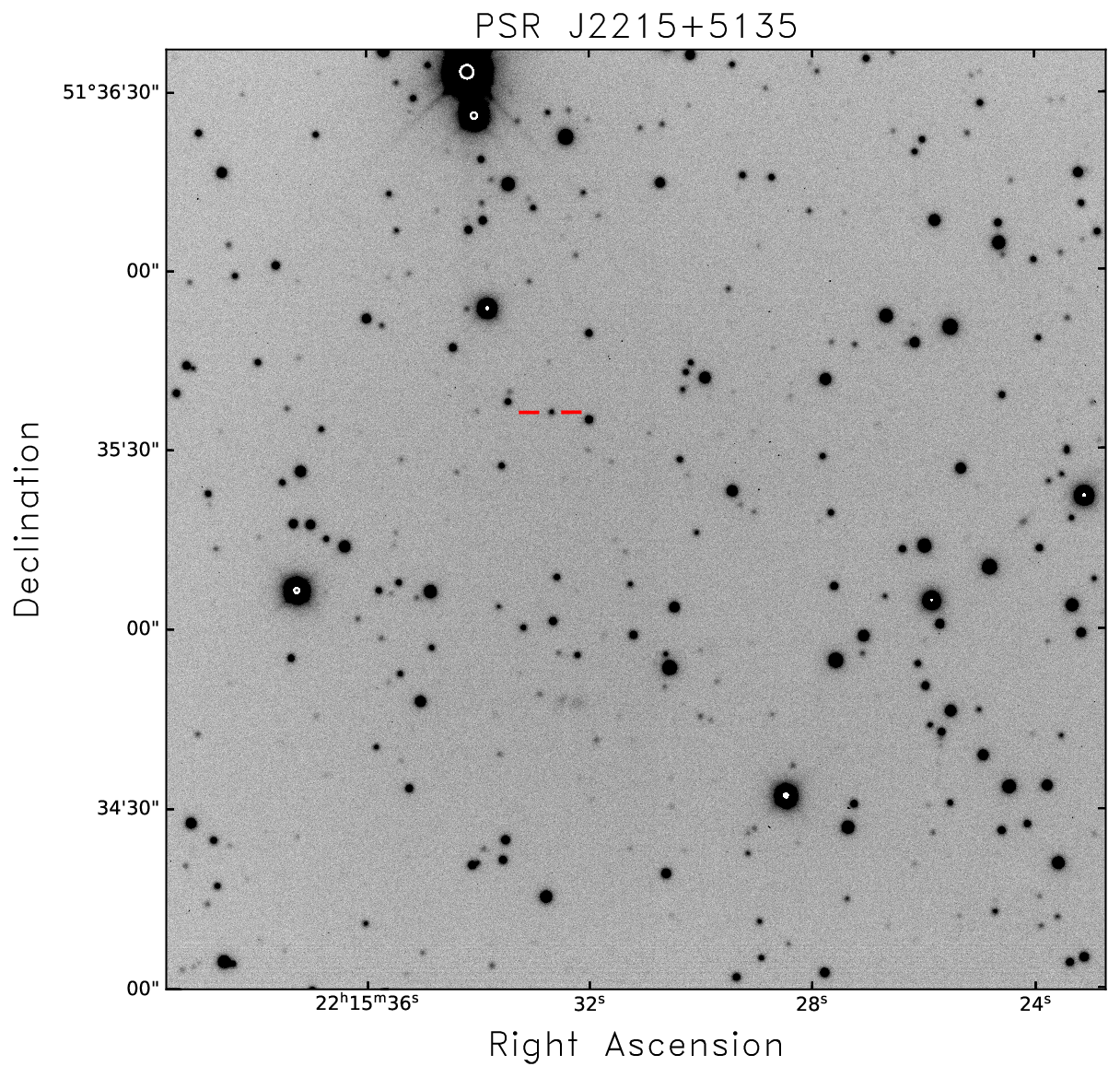}
\caption{Observation of the PSR J2215+5135 field in the \(r'\) filter over a \(3.31 \times 3.31\) arcmin\(^2\) field of view, with an exposure time of 94 s. The red dashes show the location of the redback.}
\label{fig:psr_j2215_wcs_plot}
\end{figure}

Observations were obtained with the Kepler KL4040 front-illuminated CMOS sensor camera mounted on the INO telescope. The telescope has a plate scale of 5.39 arcsec mm$^{-1}$. The detector consists of 4096 $\times$ 4096 pixels with a pixel size of 9 $\mu$m, corresponding to a pixel scale of 0.048 arcsec pixel$^{-1}$ and a total field of view of approximately 3.31 $\times$ 3.31 arcmin$^{2}$ (see Figure~\ref{fig:psr_j2215_wcs_plot}).

The observations were carried out during 2024 September 9th--11th under photometric conditions, with a typical seeing of $\sim$ 0.8 -- 1.2 arcsec. Imaging was performed through the SDSS $g'$, $r'$, and $i'$ filters, yielding a total of 237 data points across the three bands. Each filter was observed on a separate night, for a planned continuous sequence of approximately 4.14 hours. The coverage is not complete because power interruptions at the site made the telescope unavailable for part of the time, and because the target was setting toward the horizon and reached high airmass toward the end of each sequence, when we stopped observing. As a result, parts of the orbit are missing in $r'$ and $i'$. Each exposure time was 94 seconds per frame. The readout time between consecutive exposures was approximately 20 seconds.

The observing sequence was designed to ensure complete phase coverage of the PSR J2215+5135 orbital period. In particular, the observations were scheduled so that the phase $\phi = 0.5$ occurred near the midpoint of the observing sequence, corresponding to the time when the target was closest to the zenith and therefore observed at minimum airmass. As a result, phases near peak were obtained under optimal atmospheric conditions, while phases approaching $\phi = 0$ and $\phi = 1$ were located toward the higher-airmass ends of the sequence. This strategy minimizes atmospheric effects on the most critical part of the light curve. We observed one filter per night, without changing filter, by design, to avoid night-to-night differences within a single band that could arise from unpredicted variability of the system. With only three nights of observing time, we could not repeat the sequences to fill the gaps.

\bigskip
\textbf{Data Reduction:} Initial data reduction was performed using standard CCD calibration procedures. After editing the FITS headers to conform to standard keyword conventions, we constructed a master bias frame (superbias) from the set of bias exposures. This superbias was subtracted from all calibration and science frames.

Dark frames were obtained with exposure times as close as possible to the science exposures. When dark frames with different exposure times were present, they were scaled to match the science exposure time before combination. The resulting frames were median-combined to produce a master dark (superdark), which was then subtracted from all flat-field and science images.

To correct for pixel-to-pixel sensitivity variations in the CCD, twilight flat fields were obtained whenever possible. Twilight flat frames were first bias- and dark-corrected and then combined to create master flat fields. Because the CCD bias level is approximately 92 ADU, we only used flat frames with signal levels greater than 150 ADU to ensure that no negative pixels remained after bias subtraction. The individual flats were scaled before combination, and the resulting master flats were used to correct all science frames for pixel-to-pixel sensitivity variations.

Astrometric calibration was performed using \texttt{SCAMP} together with \texttt{SExtractor}, allowing us to compute accurate \texttt{WCS} solutions for each frame. Using these solutions, we cross-matched detected sources with the Pan-STARRS catalog, thereby assigning reference magnitudes to stars in the observed field.

\bigskip
\textbf{Measurements:}
Photometric measurements were carried out using the Python implementation of the \texttt{photutils} package. Both aperture photometry and PSF photometry were performed for stars within the field.

To determine the photometric zero point, we selected approximately 50 calibration stars in the vicinity of PSR J2215+5135 that have reliable magnitudes in the Pan-STARRS catalog. Instrumental magnitudes derived from our observations were compared with catalog magnitudes to determine the zero point for each filter. The Pan-STARRS1 magnitudes were converted to the SDSS system following the transformations of \cite{2012ApJ...750...99T}, because our filters follow the SDSS $g'r'i'$ system.

A comparison between our measured magnitudes and the Pan-STARRS catalog values shows that PSF photometry provides the best agreement, with a typical scatter of $\sim$ 0.02 mag between the two datasets. This agreement demonstrates that the photometric calibration of our observations is consistent with the Pan-STARRS system.

After determining the zero point, we computed the magnitudes of PSR J2215+5135 in all frames. Finally, the magnitudes were corrected for atmospheric extinction using the measured airmass and standard atmospheric extinction coefficients, as well as correcting for nonlinearity. Because the field of view is relatively small, color-dependent extinction and higher-order extinction terms are expected to be negligible; however, we applied these corrections to further minimize residual differences between our calibrated magnitudes and the Pan-STARRS reference values.


\section{Modeling Optical Data}
\label{sec:data}
To determine the binary parameters of PSR J2215+5135, we modeled the multi-band optical light curves using the Eclipsing Light Curve (ELC; Version 7; \citealt{orosz2000use}). The code models interacting binaries by combining Roche geometry with realistic stellar atmosphere models, accounting for tidal distortion, gravity darkening, limb darkening, pulsar irradiation, and surface inhomogeneities. The companion star was assumed to fill almost its Roche lobe and to rotate synchronously in a circular orbit around the neutron star. Surface intensities were computed using the NextGen atmosphere models \citep{Hauschildt1999}, with Kurucz atmospheres adopted for temperatures above 12,000 K. Irradiation from the pulsar was modeled as isotropic X-ray heating characterized by the luminosity $L_{\rm X}$, which modifies the local effective temperature of the illuminated hemisphere.

In ELC, the binary geometry is described by the orbital inclination $i$, the mass ratio $Q = M_2/M_1 $ (where $M_2$ and $M_1$ are the masses of the companion and the neutron star, respectively), Roche-lobe filling factor $f$, and orbital period $P_{\rm b}$. Several parameters were fixed from radio timing measurements, including $P_{\rm b}$, the projected semi-major axis $x$, and the pulsar mass function. We adopted synchronous rotation ($\omega=1$), a bolometric gravity-darkening exponent $\beta=0.1$ (such that $T_{\rm eff}\propto g^{\beta}$; \citealt{2021RNAAS...5..256T}), appropriate for a convective envelope, and a pulsar luminosity of $L_{}\sim10^{34}\,\mathrm{erg\,s^{-1}}$. The spin-down power of $\dot{E}=5.29\times10^{34}$\,erg\,s$^{-1}$ ($\log_{10}\dot{E}=34.72$) \citet{2013ApJ...769..108B}, is the maximum power available for heating. The posterior is $\log_{10} L_{\rm X}=34.70^{+0.10}_{-0.05}$, corresponding to a heating efficiency $L_{\rm X}/\dot{E}=0.95^{+0.24}_{-0.10}$, consistent with unity. The best-fitting solution was obtained using the Markov Chain Monte Carlo (MCMC) optimizer implemented in ELC. Flat priors were adopted for all free parameters unless otherwise specified, and the posterior distributions were used to determine the best-fitting parameter values and their corresponding uncertainties. Table \ref{tab:comparison} presents our best-fit parameters and their (1$\sigma$) uncertainties, together with a comparison to a previous study of PSR J2215+5135 (see appendix Fig.~\ref{fig:corner}).

\begin{table*}[htbp]
\centering
\caption{ELC model parameters for PSR~J2215+5135 across epochs.}
\label{tab:comparison}
\vspace{4pt}
\small

\begin{tabular}{lcccc}
\toprule
Parameter & Linares (2018)$^{a}$ & Linares (ELC) & Linares+spot (ELC)$^{b}$ & INO+spot \\
\midrule
$f$                     & $0.95 \pm 0.01$ & $0.86 \pm 0.01$ & $0.89 \pm 0.01$ & $0.96 \pm 0.01$ \\
$q$                     & $0.144 \pm 0.002$ & $0.142 \pm 0.005$ & $0.141 \pm 0.005$ & $0.144 \pm 0.009$ \\
$i$ ($^{\circ}$)        & $63.9^{+2.4}_{-2.7}$ & $64.36^{+1.31}_{-0.02}$ & $65.07^{+2.06}_{-0.14}$ & $68.05 \pm 1.26$ \\
$K_2$ (km s$^{-1}$)    & $412.3 \pm 5.0$ & $404.6^{+6.0}_{-0.7}$ & $399.9^{+10.5}_{-1.3}$ & $404.3^{+5.1}_{-5.4}$ \\
$T_\mathrm{eff,2}$ (K)    & $5630^{+52}_{-71}$ & $6590^{+55}_{-88}$ & $6075^{+181}_{-139}$ & $5404^{+177}_{-93}$ \\
$M_1$ ($M_\odot$)       & $2.27^{+0.17}_{-0.15}$ & $2.12^{+0.15}_{-0.13}$ & $2.13^{+0.16}_{-0.13}$ & $1.95^{+0.11}_{-0.07}$ \\
$M_2$ ($M_\odot$)       & $0.33 \pm 0.02$ & $0.30 \pm 0.01$ & $0.30 \pm 0.01$ & $0.28 \pm 0.01$ \\
Spot lat\ ($^{\circ}$)$^c$ & ... & ... & $1.00^{+2.76}_{-0.00}$ & $3.1^{+14.2}_{-3.0}$ \\
Spot lon\ ($^{\circ}$) & ... & ... & $263.0^{+43.4}_{-1.4}$ & $359.7^{+1.8}_{-11.2}$ \\
Spot radius\ ($^{\circ}$) & ... & ... & $14.99^{+1.8}_{-2.2}$ & $15.71^{+1.8}_{-1.4}$ \\

Spot T\ (K) & ... & ... & $9992.5 \pm 140$ & $8934.2 \pm 100$ \\

\midrule
$\chi^2$/dof & $164/126=1.30$ & $436/337=1.29$ & $423/333=1.27$ & $339/275=1.23$ \\
\bottomrule
\end{tabular}

\vspace{4pt}
\begin{minipage}{\linewidth}
\footnotesize Best-fit parameters and $1\sigma$ uncertainties from ELC modelling of the 2014 data \cite{linares2018peering} (with and without a hot spot) and the 2024 INO data (with a hot spot). The Linares (2018) column gives the published values for comparison. $f$ = Roche-lobe filling factor; $q = M_2/M_1$; $i$ = inclination; $K_2$ = companion radial-velocity semi-amplitude; $M_1$ and $M_2$ = neutron-star and companion masses; $T_\mathrm{eff,2}$ = companion effective temperature. Spot parameters are in degrees; spot temperature in K.

\footnotesize $^{a}$Published values from \cite{linares2018peering} for reference. \\
\footnotesize $^{b}$``Linares+spot (ELC)'' denotes the current work's ELC re-fit of the 2014 data including a hot spot.\\
\footnotesize $^{c}$Spot latitude is measured from the companion's north pole ($90^{\circ}$ is the orbital plane) and longitude from the inner Lagrangian point L1 ($0^{\circ}$ at L1), as in ELC.\\
\footnotesize In 2014, the spot latitude ($1^{\circ}$, measured from the pole) is far smaller than its angular radius ($15^{\circ}$), so the pole lies inside the spot and the spot longitude is effectively unconstrained.
\end{minipage}
\end{table*}

\begin{figure}[ht!]
\includegraphics[width=\linewidth]{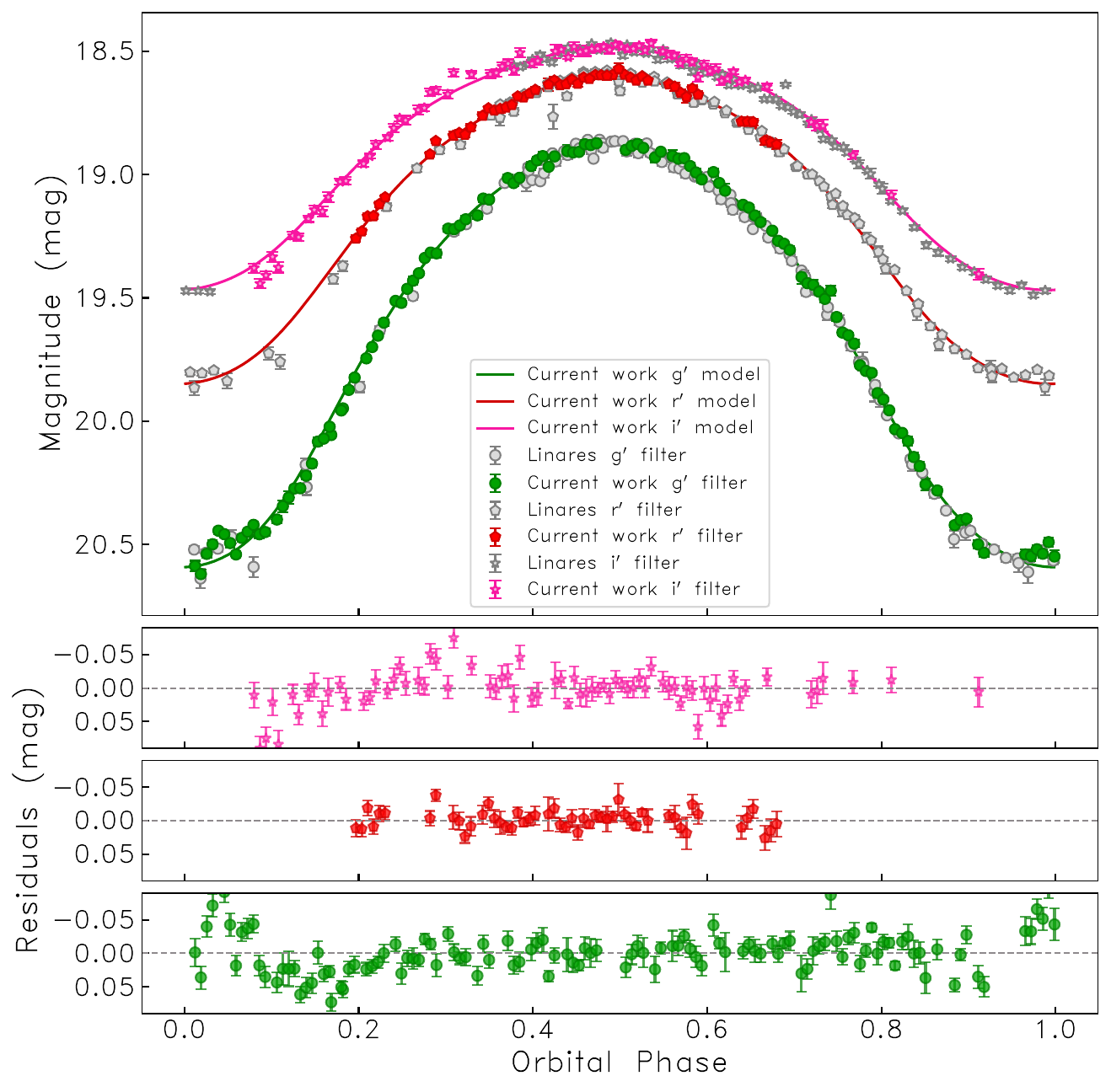}
\caption{Comparison of the 2014 \cite{linares2018peering} data with the 2024 INO observations. Gray points denote the \citeauthor{linares2018peering} data, while the green circles, red pentagons, and magenta stars represent the INO \(g'\), \(r'\), and \(i'\) data, respectively. The solid curves show the best-fitting ELC model (to the INO data). The lower panel displays the residuals between the INO observations and the best-fit model. \(i'\) data are offset upward by 0.13 mag for clarity.}
\label{fig:obs_mag_data}
\end{figure}

Orbital phases for both epochs were computed from a single ephemeris. The epoch $T_0$ = 56976.9501 $\pm$ 0.0003 (stat) $\pm$ 0.0008 (syst) MJD (TDB) comes from radial-velocity spectroscopy in \cite{linares2018peering}. It defines orbital phase $\phi = 0$, at which the companion presents its night side to the observer. The orbital period $P_b$ = 0.1725021049(8) d comes from precise radio-pulsar timing by \cite{abdo2013second}. This ephemeris has a phase-zero-point uncertainty of $\sigma_{T_0}/P_b \approx 0.005$ cycles, set by the quoted $T_0$ uncertainty. Because the same $T_0$ is used to fold both epochs, its uncertainty shifts the absolute phase of both light curves equally and does not affect their relative phase. The uncertainty from $P_b$ can affect the relative phase, but its contribution is negligible ($\lesssim 10^{-4}$ cycles) even over the $\sim$10-yr baseline to the INO epoch. A radio timing ephemeris covering the 2024 epoch is not available to us, so the registration is also tested directly through the peak comparison below, which sets the final registration uncertainty. Because orbital phase depends directly on absolute time, an uncorrected timing error translates into a corresponding phase shift. We therefore corrected the INO timestamps for a local control-system clock offset, verified against GPS time, prior to folding. All time conversions were done with the \texttt{astropy.time} package. The local time was converted to UTC for the mid-exposure time. The barycentric light-travel time was then computed for the position of the target as seen from the INO site, using the built-in astropy solar-system ephemeris, and added to the TDB time. This yields the barycentric MJD (TDB) of each exposure. The $T_0$ of Linares et al. (2018) is also given in the same time system. No empirical adjustment is applied to either light curve.

Then, following \cite{linares2018peering}, we first explored the parameter space and scaled the uncertainties in each filter so that the total reduced \(\chi^2\) of the fit was approximately unity for each dataset. We finally repeated the ELC fitting procedure to obtain the final set of parameters.

\section{Results}
\label{sec:results}
To investigate light-curve evolution in this strongly irradiated system, we compare the 2024 INO data with the 2014 measurements of \cite{linares2018peering}. For a consistent and unbiased comparison between epochs, we refit the 2014 data using the same hot-spot-inclusive model adopted for the 2024 analysis, rather than relying on the original hot-spot-free fit. This approach is motivated by the fact that omitting a physically relevant component does not merely leave residual structure; in a \(\chi^2\)-based fit, the unmodeled signal can be absorbed by other free parameters, thereby biasing the inferred solution. A direct comparison between a hot-spot model in 2024 and a hot-spot-free model in 2014 would also mix genuine secular evolution with systematic effects from model misspecification, making the two contributions difficult to disentangle.

Figure~\ref{fig:obs_mag_data} shows the two-epoch comparison. The 2024 light curves reproduce the 2014 observations in peak brightness, phase of maximum, and overall amplitude, with the magnitude at maximum light unchanged to within the measurement uncertainties in all three bands. The solid curves show the best-fit ELC model to the 2024 INO data, while the gray symbols show the 2014 WHT data in each filter.

\begin{figure*}
\includegraphics[width=0.98\linewidth]{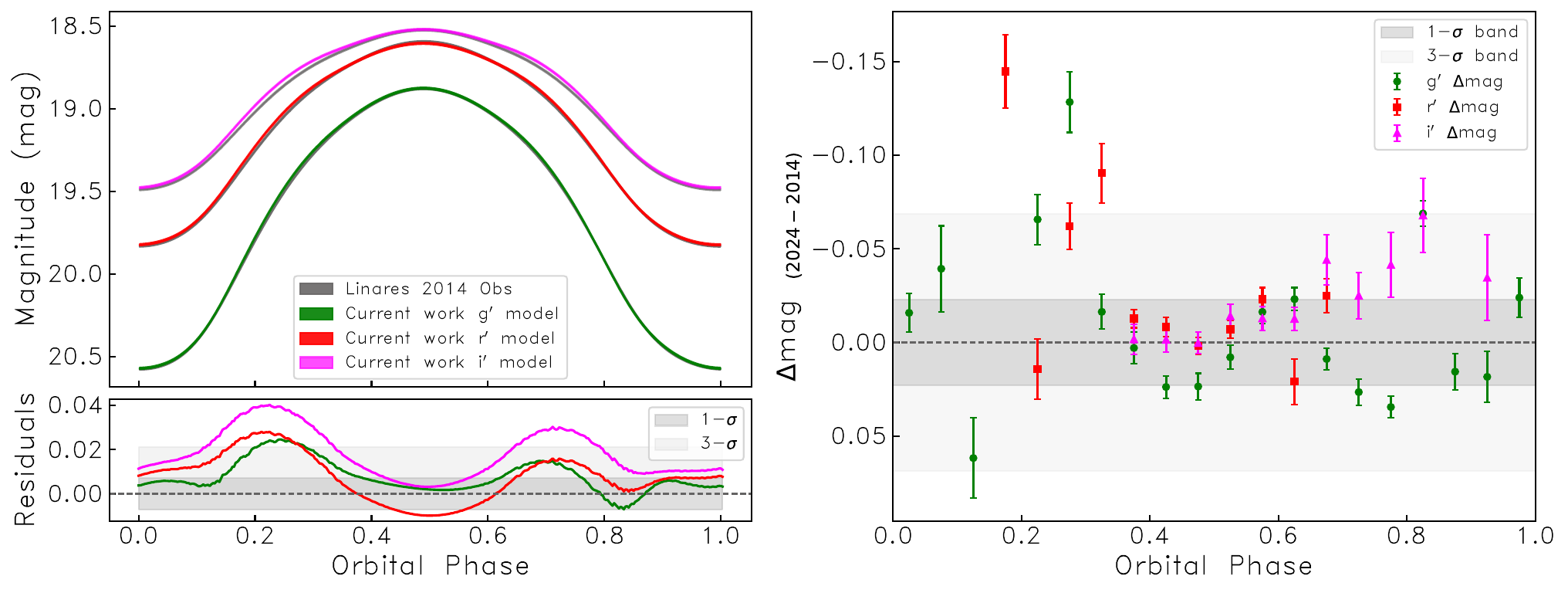}
\caption{\textit{Left panel:} Comparison of the 2014 and 2024 ELC hotspot models. The gray curves show the 2014 model, while the colored solid curves show the 2024 model. For clarity, the \(i'\) band has been shifted upward by 0.1 mag in both models. The lower panel shows the residuals in each filter. The gray shaded region denotes the \(1\sigma\) uncertainty band, and the lighter gray region indicates the \(3\sigma\) range. \textit{Right panel:} Binned 2024 - 2014 data differences shown as scatter points with vertical error bars representing the propagated \(1\sigma\) uncertainty of the binned epoch difference.}
\label{fig:mag_model}
\end{figure*}

However, a comparison of the best-fit ELC models at the two epochs (Figure~\ref{fig:mag_model}) reveals that the light-curve wings, that is, the orbital phases on either side of the peak, are noticeably broader in 2024 than in 2014. The most pronounced difference is on the rising branch (phases $0.12$--$0.33$), where all three bands show a significant flux excess relative to the 2014 data (Fig.~\ref{fig:mag_model}, left panel). The $i'$ band also shows a significant excess on the falling branch, whereas the corresponding $g'$ and $r'$ excesses on that side remain below the $3\sigma$ threshold.

To confirm that this signal is intrinsic to the data rather than an artefact of the model fit, we performed a model-independent comparison by binning both datasets into identical phase intervals and directly comparing the observed values. As shown in the right panel of Fig.~\ref{fig:mag_model}, the \(g'\) and \(r'\) bands exhibit clear deviations from the zero-residual line over the same phase range where the model comparison indicates discrepancies, supporting the conclusion that the effect is intrinsic to the data rather than an artifact of the model fit. In the \(i'\) band, a significant deviation appears on the right wing, while the left wing lacks sufficient coverage for an equally robust assessment. We note that this does not undermine the significance of the $i'$-band deviation reported in the model comparison; ELC modeling fits all three bands jointly with radial velocity constraints, so even a modest number of data points in this region carries meaningful weight in the global fit, a sensitivity that per-bin direct comparison cannot recover. Outside the discrepant interval, the binned data are consistent with zero within the uncertainties, indicating stable behavior over the remainder of the orbit.

To assess the significance of this deviation, we examine potential systematic effects. A primary concern is contamination in the \(i'\) band from stray or scattered light at the INO site. We mitigate this by excluding known infrared light sources during observations and by inspecting flat-field residuals for signatures of scattered illumination. Because the telescope rotates during the observations, any external light source would produce rotating or spatially varying patterns in the flat fields rather than fixed features on the detector, particularly in the \(i'\) band, which is sensitive to scattered light. We detect no such patterns. The root-mean-square of the flat-field residuals is \(\lesssim 0.5\%\) across the detector, corresponding to an upper limit of \(\lesssim 0.005\) mag on any systematic contribution. This is well below the amplitude of the observed deviation, indicating that scattered light is unlikely to account for the effect.

We further assess the photometric calibration by comparing more than 2,400 field stars with Pan-STARRS standards, finding consistent magnitudes across the full sample. Both aperture and PSF photometry yield a dispersion of \(\sigma \approx 0.03\) mag relative to the catalog values (see appendix Fig.~\ref{fig:mag_comp}). Although this is comparable to the amplitude of the observed deviation, two points argue against a calibration origin. First, this dispersion reflects a global uncertainty across all stars and frames, whereas the band excess is phase-dependent and confined to specific orbital intervals. Second, this scatter enters through the zero-point determination and is already incorporated into the photometric uncertainties used in the ELC modeling. We therefore conclude that the observed deviation is unlikely to arise from instrumental or calibration systematics.

Because each filter was observed on a separate night, the $g'$, $r'$, and $i'$ light curves correspond to three independent epochs rather than a single combined dataset. The phase-localized deviations relative to the 2014 epoch are consistently detected across all three bands/night individually. A night-specific transient would not be expected to reproduce the same phase-coherent feature across three unrelated nights, so this consistency argues against a single anomalous night as the cause. We acknowledge that three consecutive nights within a single observing run cannot fully exclude a transient spanning the three-night window, though we are not aware of contemporaneous evidence (e.g., X-ray or radio monitoring) for such activity.

As an independent check, we compared the phase of the light-curve peak between the two epochs. We used only the phase range $0.42$--$0.58$, which is well away from the phase ranges where the light curves differ between epochs ($0.11$--$0.33$ and $0.65$--$0.80$). In this range, the peak has the same magnitude and shape in both epochs. Because this part of the light curve is unaffected by the changes we are studying, it gives a registration check that does not depend on the very feature under investigation. We fit the peak in each epoch with Gaussian, Lorentzian, and skewed-Gaussian profiles and estimated the uncertainty with Monte Carlo resampling of the fit covariance. The resulting peak-to-peak phase difference is $\Delta\phi_{\rm peak} = -0.0007 \pm 0.0527$ cycles, consistent with zero. We adopt $\sigma_{\Delta\phi}=\sqrt{\sigma_{\rm peak}^2+\sigma_{P}^2}\approx 0.053$ cycles as the uncertainty on the phase registration between epochs, where $\sigma_{\rm peak}=0.0527$ cycles is the peak-fit uncertainty and $\sigma_{P}=N\sigma_{P_b}/P_b<10^{-4}$ cycles is the contribution of the period uncertainty over the $N$ orbital cycles between the epochs. The $T_0$ uncertainty is not included, because $T_0$ is the same for both epochs and affects only the absolute phase, not the phase difference. Since $\Delta\phi_{\rm peak}$ is well within $\sigma_{\Delta\phi}$, the peak phase is consistent with being unchanged between the epochs.

We then used this uncertainty to test whether a residual phase offset between the two epochs could account for the observed light-curve differences. We scanned a grid of small phase shifts and computed the binned $\chi^2$ difference between the 2024 and 2014 light curves. The best-fit shift was $\Delta\phi = -5\times10^{-4}$, consistent with the peak-registration result above. No rigid phase translation within the allowed uncertainty range can remove the discrepancy, confirming that the observed difference reflects a real change in light-curve morphology rather than a timing artifact.

To account for the look-elsewhere effect inherent in our multi-band, multi-phase search, we applied a trials correction by computing the global significance of the strongest deviation using a permutation test. Specifically, we asked if there were no real difference between the two light curves, what is the probability that random noise alone would produce a feature at least as strong as the observed bump anywhere in the searched region (all bands, phase wings, and window widths)? Under our primary null hypothesis in which standardized residuals are randomly shuffled in phase, the global p-value is \(p = 2 \times 10^{-5}\), corresponding to a global significance of \(4.26\sigma\). As cross-checks, we considered two additional null models: a residual bootstrap that resamples residuals with replacement, yielding \(3.76\sigma\), and a block bootstrap with block size 3 that allows for phase-correlated structure, yielding \(\sim 2.63\sigma\). These results indicate that the observed bump is highly unlikely to arise from random noise under an independent-noise model, and remains significant even when allowing for correlated residuals via a block bootstrap. The detection is therefore robust across a range of assumptions about residual structure (Appendix tab.~\ref{tab:trials}).

We next ask whether a point-source heating model with a single hot spot can account for the change. The model reproduces the broader wings in 2024 only through a combination of a changed spot and shifts in the inclination, the filling factor, and the companion temperature (Table~\ref{tab:comparison}), and it leaves systematic residuals. For the best-fit parameters, the spot remains entirely visible throughout the orbit, satisfying \(i+\mathrm{colat}+r<90^\circ\)\footnote{An approximate condition that ignores limb darkening and Roche-lobe distortion}. The corresponding sums are \(81.06^\circ\) in 2014 (\(65.07^\circ+1.0^\circ+14.99^\circ\)) and \(86.86^\circ\) in 2024 (\(68.05^\circ+3.1^\circ+15.71^\circ\)). Thus, the 2024 solution lies particularly close to the visibility limit. At the upper \(1\sigma\) value of the spot colatitude (\(17.3^\circ\)), the sum increases to \(101.06^\circ\), meaning that part of the spot could become hidden during some orbital phases. In both epochs, the rotation pole lies within the spot, making it nearly symmetric around the rotation axis. It therefore contributes an additional flux that changes little with orbital phase, increasing the brightness away from the light-curve maximum. To preserve the observed peak brightness, the fit appears to compensate by lowering \(T_{\rm eff,2}\) and adjusting \(i\) and \(f\) (Table~\ref{tab:comparison}). Thus, the broader wings can arise from the interplay between spot geometry and other model parameters, without requiring a compact heated region at a preferred longitude. Fixing the hot-spot parameters at their 2014 best-fit values and refitting the 2024 data increases the reduced chi-square from 1.23 to 1.36, while removing the hot spot increases it to 2.36. Within our adopted model, the 2024 data are therefore better fitted when a hot spot is included and its parameters are allowed to differ from their 2014 values. However, a compact spot should not be interpreted as a direct measurement of a localized heating region: its fitted contribution is strongly coupled to the spot geometry and other binary parameters. The spot is better regarded as an effective description of the additional heating, for which extended and variable heating provides a plausible physical alternative. We have only two observing epochs, 2014 and 2024, so the spot parameters can be compared at two points in time. For the 2014 data we have two solutions, namely (i) our ELC fit with point-source heating and (ii) the fit of \citet{2020ApJ...892..101K}, who analysed the same 2014 data with the ICARUS code. Comparing these two fits shows how much the spot parameters depend on the model when the data are held fixed. For the same 2014 data, our ELC fit gives a spot temperature of 9992.5\,K, compared with 9091.2\,K in the ICARUS fit, a difference of about 900\,K. The fitted companion temperature also depends strongly on the model. For the 2014 data, $T_{\rm eff,2}$ is $5630$\,K in the published fit of \citet{linares2018peering}, $6590$\,K in our spot-free ELC refit and $6075$\,K with a spot, and $f$ is $0.95$, $0.86$ and $0.89$, respectively (Table~\ref{tab:comparison}). The like-for-like change between epochs with the same spot model ($6075$\,K to $5404$\,K, about $670$\,K) is therefore smaller than the model dependence (about $960$\,K between the published and refitted values), and the $T_{\rm eff,2}$ trend is not evidence of physical cooling of the companion. Between our own ELC fits to the 2014 and 2024 data, the temperature changed from 9992.5\,K to 8934.2\,K (about 1,060\,K). For the 2014 data, the hot-spot model lowers $\chi^2$ by only 13 for four additional parameters ($\Delta$BIC $>$ 0), so the data do not require a hot spot at that epoch, and the 2014 mass is the same with and without it (2.12 vs 2.13 M$_\odot$). We retain the 2014 spot solution only so that both epochs are fitted with the same model; its parameters should not be interpreted physically. The spot colatitude ($1.0^{\circ}$ and $3.1^{\circ}$) and radius ($14.99^{\circ}$ and $15.71^{\circ}$) are similar in the two epochs, but the colatitude is poorly constrained near the pole. \citeauthor{2020ApJ...892..101K} find $16.8^{\circ +5}_{-8.4}$ and $15.5^{\circ +3.9}_{-6.8}$, respectively, so the colatitude is also model-dependent. We therefore do not claim a systematic drift. We therefore do not claim a systematic drift. The data show that the best-fit spot differs between the two epochs by an amount comparable to its dependence on the model. Whether the lower spot temperature in 2024 explains the band dependence of the excess has not been tested here, so we do not treat it as an established explanation.

For the 2014 data, the wind-heating model of \citet{2020ApJ...892..101K} gives a reduced chi-square of 1.43, compared with 1.28 for their hot-spot model, so extended heating is not preferred by the fits at that epoch. We have not tested it on the 2024 data.

\section{Discussion}
\label{sec:discussion}
The comparison between the 2014 and 2024 datasets reveals a clear evolution in the companion's light-curve morphology. The change is most pronounced on the left wing of all bands, while the \(i'\) band also shows a significant enhancement on the right wing.

The observed light-curve broadening may be associated with an evolution in the companion's surface temperature distribution. The invariance of the peak brightness and its phase argues against a global timing error (Sect.~\ref{sec:results}). The fitted binary parameters, however, are not independent of the heating prescription. The 2024 solution shifts $i$ from $65.1^{\circ}$ to $68.1^{\circ}$ and $f$ from $0.89$ to $0.96$ (large compared with the formal errors), together with $T_{\rm eff,2}$ and the spot parameters. We interpret these shifts as the model compensating for the broader wings, not as evidence that the orbital geometry has changed. In standard light-curve modeling, including ELC-based approaches, the irradiation pattern is assumed to be axisymmetric, with the temperature depending primarily on angular distance from the substellar point; under this assumption, the heated-face contribution is expected to be symmetric about superior conjunction. The observed asymmetric wing excess is therefore a departure from this idealized picture. One possible explanation is that energy deposition extends over a broader longitudinal range on the companion's surface, leaving the peak phase fixed while broadening the wings. The present data cannot establish this interpretation, because the near-polar effective spot in our point-source model can mimic such a component. This possibility suggests that the companion may be more than a passively heated target: its own magnetic field, atmospheric winds, and mass-loss flow could influence where and how heating occurs and how it evolves over time. A consequence is that the parameters of a point-source model with one spot, including the neutron-star mass, are conditional on the adopted heating prescription. The inferred mass should therefore be interpreted within the context of this model and its assumptions.

The multiband character of the signal helps constrain the nature of the change. The flux excess is strongest in $i'$ and weakens progressively toward bluer wavelengths. Whether this wavelength dependence is related to the lower best-fit spot temperature in 2024 has not been tested. The change in $T_{\rm eff,2}$ between epochs ($\approx 670$\,K for the same spot model) is smaller than its model dependence ($\approx 960$\,K between our refit and the published value), and the change in the spot temperature ($\approx 1060$\,K) is comparable to it. In addition, the thermal components of the model vary in a correlated manner, apparently driven by the requirement to keep the peak flux fixed, along with changes in \(i\) and \(f\). We therefore do not interpret these trends physically. The data do not demonstrate that the heated and unheated components changed in a physically coordinated way. The excesses in the left and right wings may arise from the same change in the companion's surface-heating pattern viewed at different orbital phases. Further multi-epoch, multiwavelength observations and consistent modeling are needed to investigate this possibility.

Several mechanisms could in principle produce such a change in the heating pattern, including intrabinary shock displacement, magnetically channelled heating at preferred longitudes, and atmospheric heat redistribution. However, these remain qualitative proposals motivated by the observed morphology rather than conclusions drawn from a physical model. Resolving the origin of this evolution will require dedicated theoretical modelling and a multi-component surface prescription that goes beyond the single axisymmetric spot assumed in current light-curve codes.

The physical origin and the timescale of the change therefore remain uncertain. Additional multi-epoch observations, regularly spaced over months and years and in several bands, are needed to determine whether the change is gradual, episodic, or more complex. Redbacks are known to vary on timescales from minutes to months, and PSR~J2215+5135, although less variable than some other redbacks, may not be stable at all times. With two epochs ten years apart, we cannot tell whether the change is a slow drift, a sudden change, a short-lived state at the time of our observations, or the effect of more than one spot. The three filters were observed on separate nights and show the same phase-localised excess in the ranges they cover, which argues against a transient lasting less than one night, but cannot exclude one lasting a few days. Regular monitoring over months and years in several bands is needed to separate these possibilities.

The hot-spot treatment has a direct and quantifiable consequence for the neutron-star mass inferred from light-curve modeling. When we fit INO 2024 photometry together with the radial-velocity data of \citet{linares2018peering}, but hold the spot geometry fixed at the values reported by \citet{2020ApJ...892..101K} from ICARUS fits to the 2014 data, we obtain $M_\mathrm{NS} = 2.14\,M_\odot$. Allowing the spot radius to vary freely brings this down to $2.0\,M_\odot$, and freeing the spot longitude and latitude as well gives $M_\mathrm{NS} = 1.95^{+0.11}_{-0.07}\,M_\odot$. The $\sim\!0.2\,M_\odot$ difference between the most and least constrained spot treatments, arising purely from how the spot is parametrised and not from the data themselves, is comparable in magnitude to the formal uncertainties quoted in individual studies, and accounts for a substantial fraction of the full spread of mass estimates published for this source over the past decade ($1.6$--$2.45\,M_\odot$; \cite{Schroeder:2014qwa, romani2015massive, linares2018peering, 2020ApJ...892..101K, Sullivan:2024qgl}). This suggests that the hot-spot prescription is an important systematic in the mass measurement of this source. The comparison between the two epochs illustrates how the inferred neutron-star mass depends on the observed light curve and the adopted heating model. For PSR~J2215+5135, the best-fit masses with a hot spot are $2.13^{+0.16}_{-0.13}\,M_\odot$ for 2014 and $1.95^{+0.11}_{-0.07}\,M_\odot$ for 2024. These values agree within about $1\sigma$, so we do not claim a significant change in the neutron-star mass. The 2024 mass is lower together with a higher inclination ($68.1^{\circ}$ versus $65.1^{\circ}$ in 2014), as expected from the mass function, and this inclination shift is one of the compensating changes discussed above; the mass difference should therefore not be read as a physical change. We adopt the fully free-moving spot fit for 2024 as our fiducial estimate because it gives the lowest reduced chi-square (1.23, compared with 1.36 when the spot parameters are fixed at their 2014 values and 2.36 without a spot). However, the agreement between the mass estimates does not establish that the inferred mass is insensitive to the heating prescription, as the quoted uncertainties are statistical and do not capture the full uncertainty associated with model assumptions. We therefore recommend using multi-epoch optical light curves, where available, to assess the robustness of neutron-star mass measurements.

Time-domain photometric monitoring of PSR~J2215+5135 and other spider pulsars would help determine whether similar long-term light-curve changes are common and how they affect neutron-star mass measurements. The decadal change observed here may reflect changes in how pulsar-wind energy is deposited and redistributed across the companion's surface, motivating theoretical models that go beyond the single axisymmetric spot assumed in current light-curve codes. Multi-epoch, phase-resolved spectroscopy could further constrain the companion's surface temperature distribution and test whether it evolves over time, helping to distinguish between competing heating scenarios and improve the reliability of mass measurements.

\begin{acknowledgments}
A.F. thanks the Iranian National Observatory (INO) for facilitating and funding these observations. We thank the High Performance Computing Centre of the Institute for Research in Fundamental Sciences (IPM) for providing the computational resources used for the ELC modelling. We also thank Dr. Manuel Linares for his constructive comments and valuable contributions, and the anonymous referee for their thoughtful and constructive suggestions, which helped improve this manuscript. The authors used Claude (Anthropic) for text editing. All scientific content and conclusions remain the sole responsibility of the authors.

\end{acknowledgments}

\facilities{INO (KL4040 camera), WHT}

\bibliography{sample701}{}
\bibliographystyle{aasjournalv7}

\appendix

\section{Additional Figures}

\begin{figure}[ht!]
\centering
\includegraphics[width=0.8\linewidth]{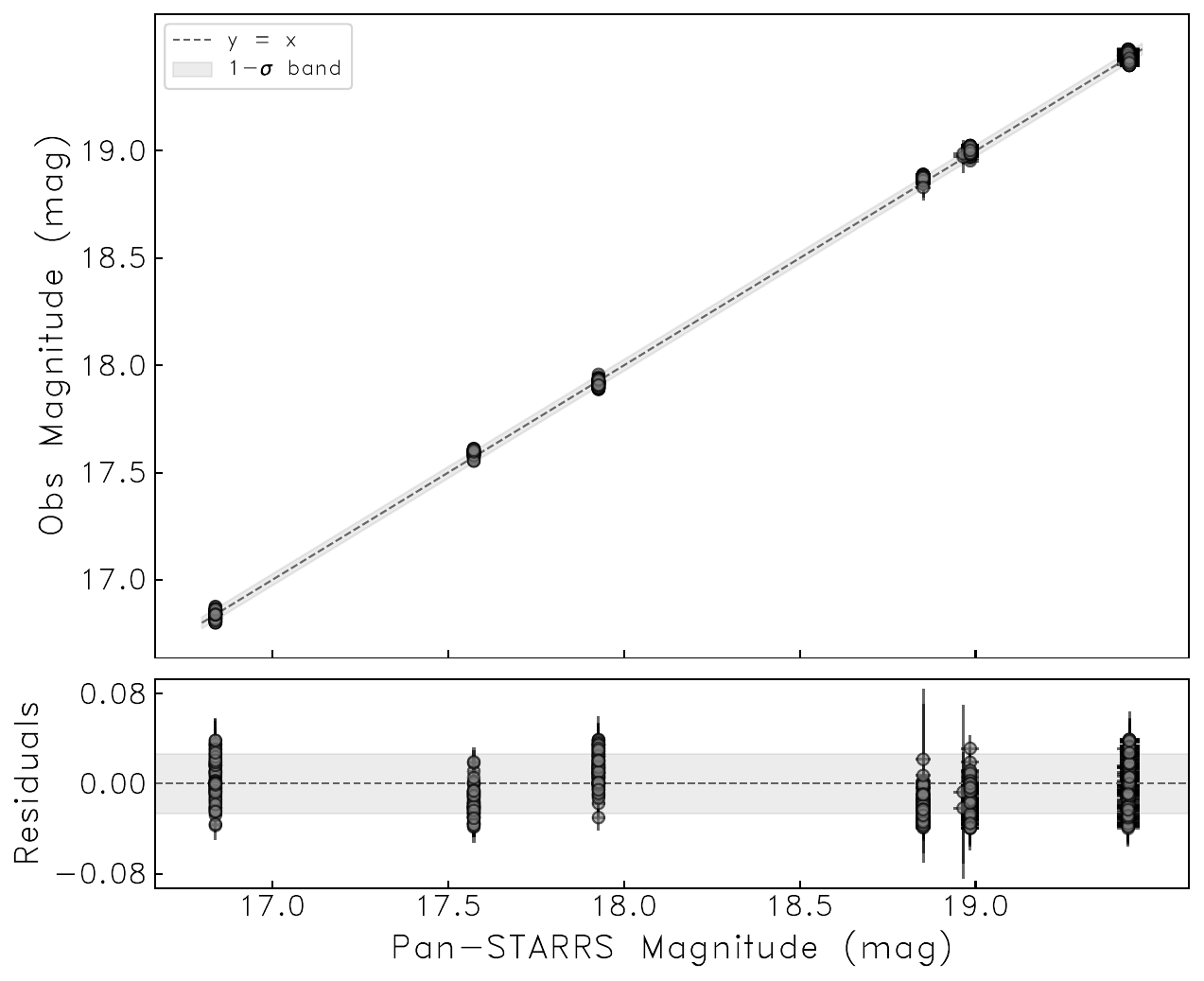}
\caption{Photometric calibration diagnostics. \textit{(Top panel)} $g'$-band PSF magnitudes measured at INO versus the Pan-STARRS catalog magnitudes for calibration stars in the PSR~J2215+5135 field. \textit{(Bottom panel)} Residuals between the measured PSF magnitudes and the Pan-STARRS catalog values. The rms scatter is $\approx0.025$\,mag, indicating a stable photometric zero point across the field.}
\label{fig:mag_comp}
\end{figure}

\begin{figure}[ht!]
\centering
\includegraphics[width=0.8\linewidth]{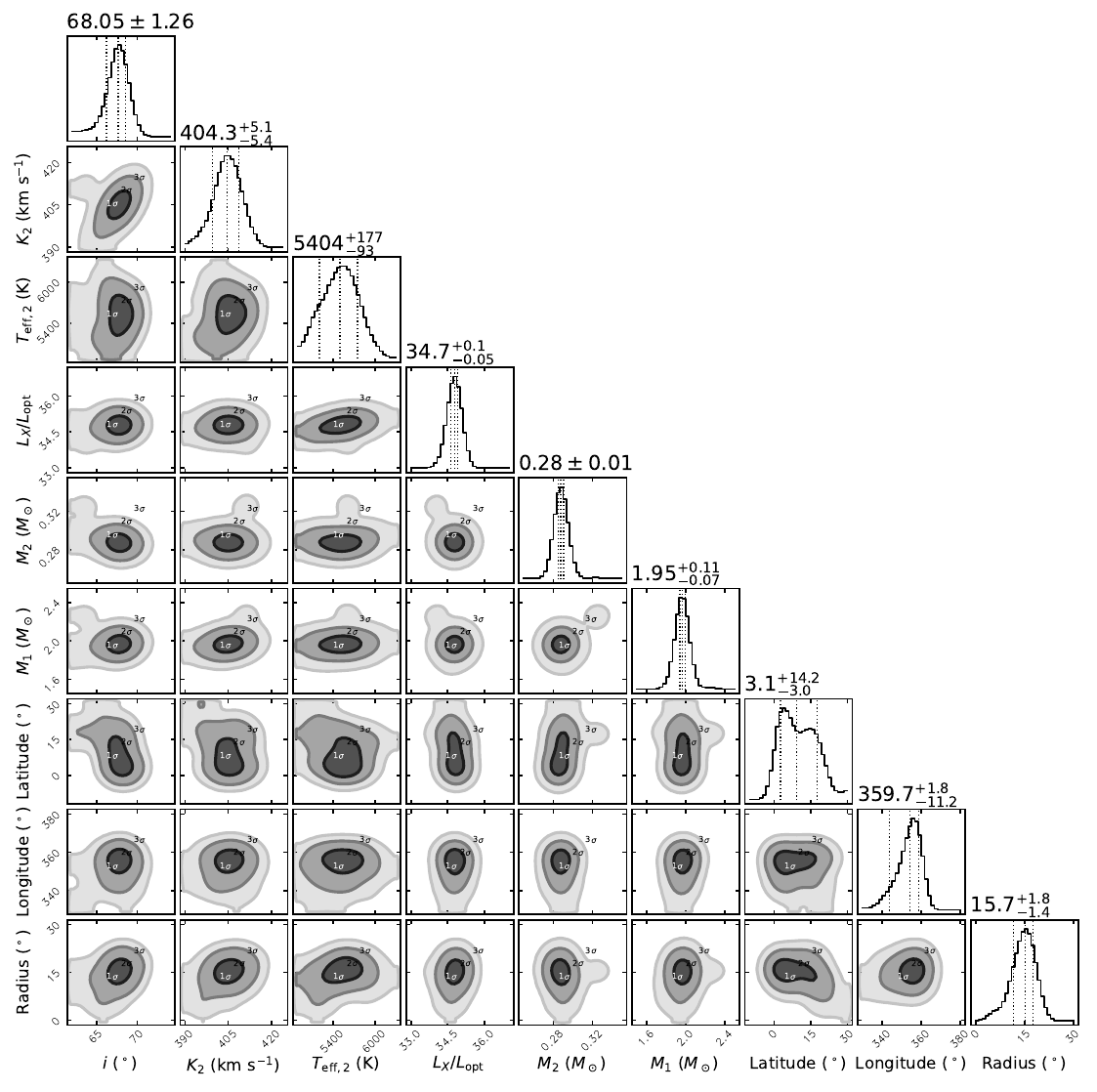}
\caption{Corner plot of the posterior distributions from the ELC model with free hot-spot parameters. The median values and $1\sigma$ uncertainties are listed at the top of each column. The contours correspond to the 1, 2, and $3\sigma$ confidence regions derived from the MCMC sampling. The corresponding numerical values are listed in Table~\ref{tab:comparison}. The $L_X/L_{\rm opt}$ is the base-10 logarithm of the X-ray luminosity, in erg s$^{-1}$.}
\label{fig:corner}
\end{figure}

\noindent
\begin{table}[htbp]
\centering
\caption{Permutation significance results.} 
\label{tab:trials}
\begin{tabular}{lcc}
\toprule
Null model & Global $p$-value & Significance \\
\midrule
Phase permutation (primary)  & $2 \times 10^{-5}$ & $4.26\sigma$ \\
Residual bootstrap           & $8 \times 10^{-5}$ & $3.76\sigma$ \\
Block bootstrap (size 3)     & $4 \times 10^{-3}$ & $2.63\sigma$ \\
\bottomrule
\end{tabular}

\footnotesize Global significance of the observed light-curve broadening under three null models. In each case, we ask what fraction of $10^5$ realisations of noise produce an excess at least as large as observed anywhere in the searched region (both wings, all bands, all window widths)? The block bootstrap (block size 3) preserves phase-correlated residual structure and provides the most conservative estimate.
\end{table}

\section*{Data Availability}
The 2024 INO photometry shown in Figure~\ref{fig:obs_mag_data} (time, orbital phase, magnitude and uncertainty in $g'$, $r'$ and $i'$) is available as machine-readable ASCII tables with the online version of this article. The 2014 data are from \citet{linares2018peering}. The best-fit parameters are given in Table~\ref{tab:comparison}.

\end{document}